\documentclass[structabstract]{aa}

\usepackage{amsmath}
\usepackage{txfonts}          
\usepackage{mathrsfs}         
\usepackage{lscape}           
\usepackage{afterpage}        
\usepackage{natbib}           
\usepackage{graphicx}
\usepackage{graphics}
\usepackage{dcolumn}
\usepackage{multicol}
\usepackage{authblk}
\usepackage{tikz-cd}
\usetikzlibrary{arrows,arrows.meta}
\tikzcdset{arrow style=tikz, diagrams={>=angle 90}}

\newcolumntype{d}[1]{D{.}{.}{#1}}

\begin{document}

\title{Accurate millimetre and submillimetre rest frequencies for \\ $cis$- and $trans$-dithioformic acid, HCSSH}

\author{D.~Prudenzano\inst{1} \and J.~Laas\inst{1} \and L.~Bizzocchi\inst{1} \and V.~Lattanzi\inst{1} \and C.~Endres\inst{1} \and B.M.~Giuliano\inst{1} \and S.~Spezzano\inst{1} \and M.E.~Palumbo\inst{2} \and P.~Caselli\inst{1}}

\institute{{Center for Astrochemical Studies, 
           Max-Planck-Institut f\"ur extraterrestrische Physik,
           Gie\ss enbachstra\ss e 1, 85748 Garching (Germany);
           \email{prudenzano@mpe.mpg.de}}
           \and
           {INAF-Osservatorio Astrofisico di Catania, I-95123 Catania, Italy;
           \email{mepalumbo@oact.inaf.it}   
           }}

\titlerunning{Rest frequencies of dithioformic acid}
\authorrunning{D. Prudenzano et al.}
%

\abstract
%
{A better understanding of sulphur chemistry is needed to solve the interstellar sulphur depletion problem. A way to achieve this goal is to study new S-bearing molecules in the laboratory, obtaining accurate rest frequencies for an astronomical search. We focus on dithioformic acid, HCSSH, which is the sulphur analogue of formic acid.}
%
{The aim of this study is to provide an accurate line list of the two HCSSH $trans$ and $cis$ isomers in their electronic ground state and a comprehensive centrifugal distortion analysis with an extension of measurements in the millimetre and submillimetre range.}
%
{We studied the two isomers in the laboratory using an absorption spectrometer employing the frequency-modulation technique. The molecules were produced directly within a free-space cell by glow discharge of a gas mixture. We measured lines belonging to the electronic ground state up to 478 GHz, with a total number of 204 and 139 new rotational transitions, respectively, for $trans$ and $cis$ isomers. The final dataset also includes lines in the centimetre range available from literature.}
%
{The extension of the measurements in the mm and submm range lead to an accurate set of rotational and centrifugal distortion parameters. This allows us to predict frequencies with estimated uncertainties as low as 5 kHz at 1 mm wavelength. Hence, the new dataset provided by this study can be used for astronomical search.}   
{}

              
\maketitle


\section{Introduction} \label{sec:intro}
\indent\indent
Sulphur chemistry has attracted interest since the early 70s, when the first S-bearing molecule, CS, was detected in the interstellar medium \citep{penzias1971cs}, followed by observations of other sulphur species in the same decade, such as OCS, H$_2$S, and CH$_3$SH (\citealt{jefferts1971ocs}, \citealt{thaddeus1972h2s}, \citealt{linke1979} and references therein). 
In diffuse clouds sulphur is mainly found in the gas phase in the form of S$^+$ and has cosmic abundances of $\sim$10$^{-5}$ with respect to the H nuclei (\citealt{jenkins1987}, \citealt{savage1996}). However, to reproduce the observed values of S-bearing molecules in dark clouds (\citealt{oppenheimer1974}, \citealt{tieftrunk1994}), chemical models typically assume that sulphur is highly depleted, using an initial elemental abundance of 8$\times$10$^{-8}$ \citep{wakelamherbst2008}. A possible explanation of this problem is given by \cite{ruffle1999}, who proposed that during collapse of translucent gas, sulphur tends to freeze-out more than other elements in dark clouds, due to its predominant ionized form, S$^+$, and thus it is subject to strong electrostatic attraction to negatively charged grains.

Many studies have discussed the problem and some have suggested that most of the missing sulphur in dense regions could be trapped in the solid phase as either polysulphanes, having the formula H$_x$S$_y$ (\citealt{jimenezescobar2011}, \citealt{druard2012}), or refractory sulphur polymers, such as S$_8$ \citep{wakelamcaselli2004, wakelamcaselli2005}. 
However, the main reservoirs of solid sulphur are still unknown.    
Some laboratory studies, focussing on dust and ice analogues, propose carbon disulfide, CS$_2$, as a major sink of sulphur (\citealt{ferrante2008}, \citealt{garozzo2010}, \citealt{jimenezescobar2014}). 
Recently, CS$_2$ has been found in the comas of some comets through the identification of its emission spectra in the UV and visible region \citep{jackson2004} or in situ by ROSINA \citep{calmonte2016}.
In particular, \cite{calmonte2016} have found that the amount of CS$_2$ in the coma of 67P/Churyumov-Gerasimenko accounted for the 0.35$\%$ of the total sulphur content, resulting more abundant than thioformaldehyde, H$_2$CS (0.14$\%$), a well-known S-bearing molecule, detected for the first time in SgrB2 by \cite{sinclair1973h2cs}.
Nevertheless, carbon disulfide is still undetected in interstellar ices. 
Furthermore, in warm and shocked regions, where this compound could be transferred in the gas phase as a result of evaporation and sputtering of the grains, its detection by means of radio telescopes is unfeasible because it has no permanent dipole moment. Therefore other S-bearing molecules must be sought.  
  
Recently, many steps have been taken towards improving our understanding of interstellar sulphur chemistry, as highlighted by the detection of new S-bearing compounds, such as S$_3$, S$_4$, and ethyl mercaptan (CH$_3$CH$_2$SH) in the comas of 67P/Churyumov-Gerasimenko \citep{calmonte2016} and a tentative detection of CH$_3$CH$_2$SH in Orion KL \citep{kolesnikova2014}.
Moreover, chemical models have been significantly improved through expansion of the sulphur chemical network by the inclusion of these compounds and many other reactions and species (\citealt{woods2015}, \citealt{vidal2017}). 
Notwithstanding this progress, a number of important molecules are still not considered in the current picture of sulphur chemistry, including dithioformic acid, HCSSH. 
This species is chemically related to CS$_2$ and is the S-bearing analogue of formic acid, HCOOH; the latter was detected for the first time in SagittariusB2 (SgrB2) by \cite{zuckerman1971} and in the dark cloud L134N by \cite{irvine1990}. 
Because it is isostructural with HCOOH, HCSSH exists in two different conformations, dubbed $cis$ or $trans$, depending on the orientation of the S-H bond with respect to the C-H bond. These two conformers can be seen in Fig. 1.

\begin{figure} [h]
   \centering
    \includegraphics[width=6.5cm]{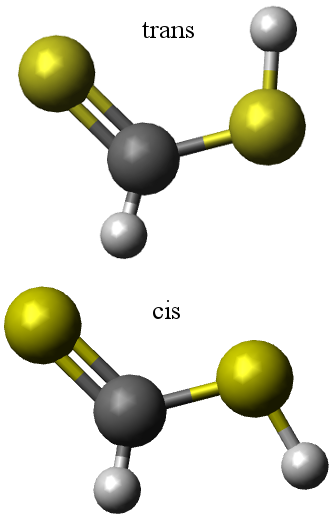}
     \caption{Molecular structures of the two planar isomers of HCSSH. Details concerning the geometries of the two species are given in Section 3 and Appendix A. Yellow spheres indicate sulphur atoms, the dark grey sphere indicates carbon, and hydrogen atoms are shown in light grey.}
      \label{isomers}
\end{figure}
 
The first laboratory study of dithioformic acid was carried out by \cite{bak1978}. They measured transitions in the centimetre (cm) regime, between 18 and 40 GHz, for the $trans$ and the $cis$ conformers in the ground and in two low-lying vibrationally excited states. They also found an abundance ratio trans/cis=5.43, corresponding to an energy difference of 350 cm$^{-1}$.
The molecules were produced by gas-phase pyrolysis of  methanetrithiol, HC(SH)$_3$, and the spectra were recorded with a Stark-modulated microwave spectrometer. 
In a subsequent work, the same authors carried out additional measurements of the two $^{34}$S and one D isotopologues, leading to the experimental determination of the $trans$-HCSSH structure \citep{bak1979}.
However, the limited frequency range covered in these investigations prevented a satisfactory spectroscopic characterization of these species. In particular, the incomplete centrifugal distortion analysis and the rather high measurement errors caused by the interfering Stark lobes\footnote{Artefacts produced by the Stark modulation close to the centre frequency.} led to high estimated uncertainties in the spectral predictions. The calculated rest frequencies are, in fact, affected by errors exceeding 0.7 MHz for the $trans$ isomer and 1.1 MHz for the $cis$, already in the 3 mm region.       

Given the potential astronomical interest of this molecule, an improved knowledge of its rotational spectrum is desirable. In this work we present a thorough centrifugal distortion analysis of both $trans$ and $cis$ isomers of HCSSH. We extended the spectroscopic measurements well into the submillimetre (submm) domain, reaching a frequency as high as 478 GHz for the $trans$-HCSSH. High-level theoretical calculations were also carried out to provide a detailed description of the molecular properties, such as dipole moment components and equilibrium structures.

\section{Experiments} \label{sec:exp}
\indent\indent
The measurements were performed using the CASAC (Center for Astrochemical Studies Absorption Cell) spectrometer at the Max-Planck-Institut f\"ur extraterrestrische Physik. Full details of the experimental apparatus can be found in \cite{bizzocchi2017}. The instrument employs the frequency 
modulation technique in the mm and submm range and is equipped with a glow discharge cell for production of unstable species. 

Measurements were carried out at room temperature to avoid condensation of the precursors. Dithioformic acid was produced by a 40\,mA DC discharge ($\sim$0.8 kV) from a 1:1 mixture of CS$_2$ and H$_2$, diluted in Ar buffer gas. 
The total pressure in the cell was 20-30 mTorr (27-40 $\mu$bar).  

\section{Results and data analysis} \label{sec:res}
\indent\indent
The HCSSH isomers are near-prolate asymmetric rotors ($\kappa_{trans}$=-0.9901 and $\kappa_{cis}$=-0.9682 \footnote{Rey asymmetry parameter. It is an indication of the asymmetry of the molecule and can take values between -1, for a prolate symmetric top, and +1, for an oblate symmetric top. \citep{gordy1984} }) with \textit{C$_s$} symmetry and planar structure. The $\textit{a}$ components of the dipole moment have been experimentally determined by \cite{bak1978}, where $\mu_{a}(trans)$=1.53 Debye (D) and $\mu_{a}(cis)$=2.10 D, while the $\textit{b}$ components were determined in our study (see Appendix A).  
We measured 204 new line frequencies for the $trans$ conformer and 139 for the $cis$. We also detected seven $b$-type transitions for the $cis$ conformer, which has a higher value of $\mu_b$, equal to 1.67 D (see Table A.2). 
These new rotational transitions belong to the $R$ branch ($\Delta J$=+1) and \textit{J} values range from 14 to 74 and a maximum $K_a$=20. 

We recovered the line central frequencies with the $\texttt{proFFit}$ code \citep{dore2003}, adopting a modulated Voigt profile.
The estimated accuracy ranges between 25--50\,kHz, depending on the line width, signal-to-noise ratio, and background continuum (standing waves produced between the two partially reflecting windows placed on either side of the cell). Examples of measured line and fitted profile of $trans$ and $cis$ isomers are shown in Figure 2. Both panels include lines corresponding to the blended $K_a$=7 asymmetry doublet, located at ca. 360 GHz. 

\begin{figure*} [th]
   \centering
    \includegraphics[width=19.cm]{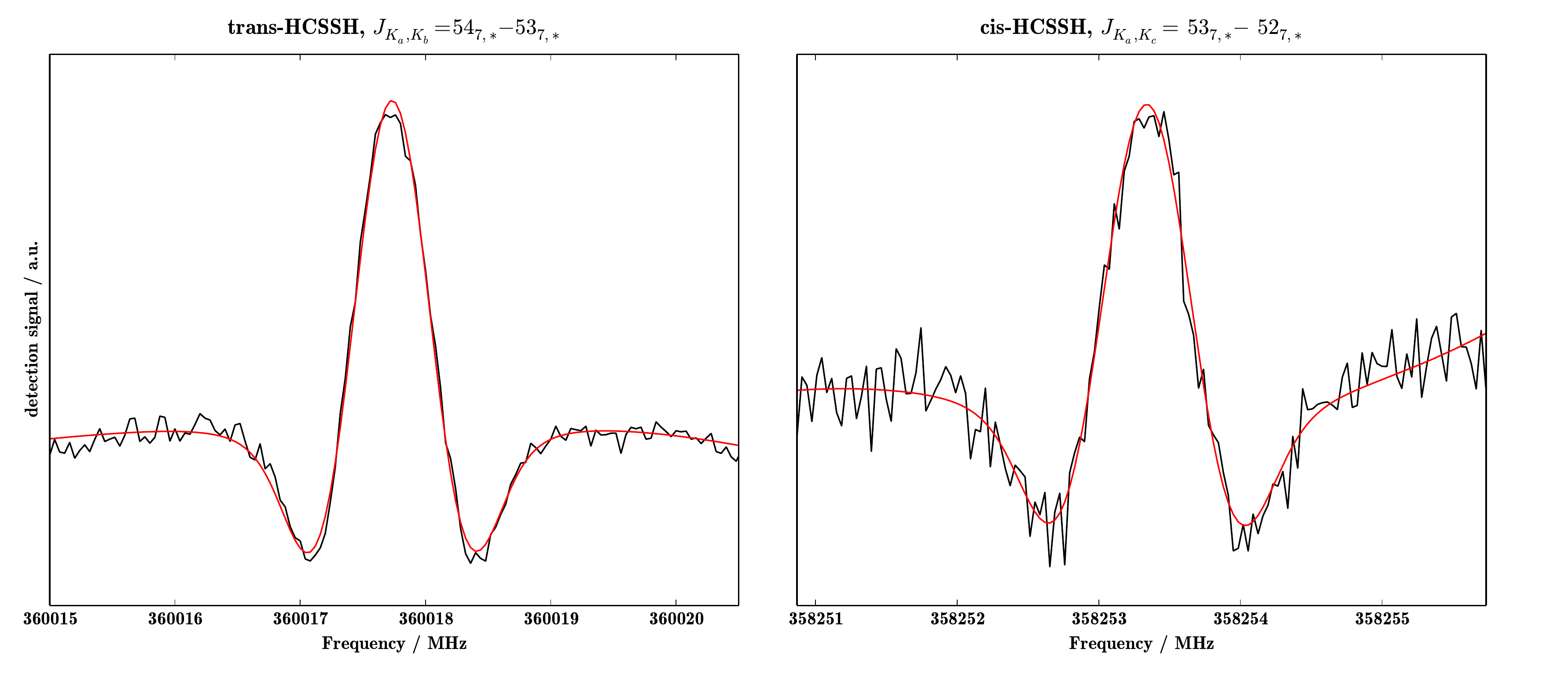}
     \caption{(\textit{Left panel}) Experimental spectra of the 54$_{7,*}$ - 53$_{7,*}$ transitions of $trans$-HCSSH, with total integration time of 146 s and 3 ms of time constant. (\textit{Right panel}) Experimental spectra of the 53$_{7,*}$ - 52$_{7,*}$ transitions of $cis$-HCSSH with total integration time of 87 s and 3 ms of time constant. (\textit{Both panels}) The $K_a$=7 asymmetry doublet is not resolved. The red traces indicate the resulting line profile fitted with the \texttt{proFFit} code (see text).}
      \label{transline}
\end{figure*}

We performed the spectral analysis using the Pickett SPFIT/SPCAT suit of programs \citep{pickett1991}, adopting the Watson S-reduced Hamiltonian for asymmetric-top molecules \citep{watson1977b}.
The transitions measured by \cite{bak1978} in the cm range were also included. 
To take into account the different experimental accuracies, we employed four distinct statistical weights (1/$\sigma^2$) for the two datasets. For the lines taken from literature, we assigned $\sigma$ values of 75 kHz and 100 kHz to the $trans$ and the $cis$ isomers, respectively.
A few lines showing very large deviation were removed from the final fit. 
Experimental uncertainties for our measurements are set to 25 or 50 kHz; the higher value is assigned only to lines broadened due to partially overlapped \textit{K} doublets. 
The final fits lead to the determination of the complete set of quartic centrifugal distortion constants and four sextic constants, i.e. $H_J$, $H_{JK}$, $H_{KJ}$, and $h_1$ (energy contributions depending on $J^6$, obtained from the rotational Hamiltonian). One octic constant, i.e. $L_{KKJ}$, was also determined for the $trans$ isomer. The weighted root mean square (RMS) deviations are 0.936 and 0.933 for $trans$ and $cis$, respectively. The resulting parameters are listed in Table 1. The lists of all the experimental frequencies are available at the CDS\footnote{http://cdsweb.u-strasbg.fr/} as supplementary data.
Planarity of the species was confirmed by calculation of the inertial defect, $\Delta =I_c - (I_b + I_a)$ \citep{darling1940}, obtained using the moments of inertia $I$, derived from the ground-state rotational constants. The inertial defect values for $trans$ and $cis$ dithioformic acid are 0.0185 and 0.0173 amu \AA$^2$, respectively, close to those of the two formic acid planar isomers, i.e. 0.0126 and 0.0102 amu \AA$^2$ (\citealt{trambarulo1958}, \citealt{davis1980} and references therein, \citealt{winnewisser2002}). 

\begin{table*} [h]
\caption{Spectroscopic parameters of the two conformers of HCSSH. The results of this study are compared with those of \cite{bak1978}.}
\label{table_all}
\begin{tabular}{l l d{-1} d{-1} d{-1} d{-1}}
\hline \hline \\ [-2.05ex]
Constants & Units & \multicolumn{1}{c}{$trans$-HCSSH\tablefootmark{a}} & \multicolumn{1}{c}{$trans$-HCSSH\tablefootmark{b}} & \multicolumn{1}{c}{$cis$-HCSSH\tablefootmark{a}} & \multicolumn{1}{c}{$cis$-HCSSH\tablefootmark{b}} \\
\hline \\ [-2ex]
\textit{A$_0$}         & MHz & 49206.11(21)$\tablefootmark{c}$  &  49227.(86)      & 48572.439(66)     & 48947.(380)
\\ [1ex]                                                           
\textit{B$_0$}         & MHz & 3447.53432(37)    &  3447.5312(89)  & 3498.74789(67)    & 3498.719(24)
\\ [1ex]                                                           
\textit{C$_0$}         & MHz & 3219.47256(37)     &  3219.4954(96)  & 3261.42278(62)    & 3261.433(30)
\\ [1ex]                                                           
\textit{D$_J$}         & kHz & 1.063389(64)      &  1.10(10)$\tablefootmark{d}$       & 1.27724(14)       & 1.02(42)$\tablefootmark{d}$
\\ [1ex]                                                           
\textit{D$_{JK}$}      & MHz & -0.0389438(19)    &  -0.03905(24)$\tablefootmark{d}$   & -0.0470807(33)    & -0.0484(12)$\tablefootmark{d}$
\\ [1ex]                                      
\textit{D$_K$}         & MHz & 1.376(33)         &                     & 1.412(17)         & 
\\ [1ex]                                                                                
\textit{d$_1$}         & Hz  & -3.609(38)        &    & 4.343(28)         & 
\\ [1ex]                                                           
\textit{d$_2$}         & kHz & -0.119787(80)     &    & 0.14900(20)       & 
\\ [1ex]       
\textit{H$_J$}         & Hz  & 0.0005714(64)     &    & 0.001169(25)      & 
\\ [1ex]
\textit{H$_{JK}$}      & Hz  & -0.00573(31)      &    & -0.03601(50)      & 
\\ [1ex]                                                                                                  
\textit{H$_{KJ}$}      & Hz  & -4.216(15)        &    & -4.109(20)        & 
\\ [1ex]                                                                                                                                   
\textit{h$_1$}         & Hz  & 0.0001673(96)     &    & 0.000288(38)     & 
\\ [1ex]                                                           
\textit{L$_{KKJ}$}     & Hz  & 0.000447(27)      &    &                   & 
\\ [1ex]
$I_c -(I_b + I_a)$ & amu $\AA ^2$ & 0.0185 &  & 0.0173 & 
\\ [1ex]
\textit{$\sigma_{w}$} & & 0.936 &  & 0.933 & 
\\ [1ex]
No. of lines & & 204 & 25 & 139 & 19
\\
\hline
\end{tabular}
\\ [-1ex]
\tablefoot{
\tablefoottext{a}{This work.}\\
\tablefoottext{b}{\cite{bak1978}.} \\
\tablefoottext{c}{Standard error in parentheses are in units of the last digit.} \\
\tablefoottext{d}{Calculated using the Watson A-reduction for asymmetric top molecules.}
}
\end{table*}

The complete line catalogues, computed by using the spectroscopic parameters in Table 1, are provided as supplementary data at the CDS and also include  the 1 $\sigma$ uncertainties, upper-state energies, line strength, and the Einstein \textit{A} coefficient for spontaneous emission, all expressed in SI units as follows:
\begin{equation}
A_{ij} = \frac{16 \pi^3 \nu^3}{3 \epsilon_0hc^3} \frac{1}{2J + 1} S_{ij} \mu ^2,
\end{equation}
where $\nu$ is the line frequency, $\epsilon_0$ is the vacuum permittivity, $h$ is the Plank's constant, $c$ is the speed of light, $J$ is the quantum number of the upper state for the rotational angular momentum, and $S_{ij} \mu ^2$ is the line strength. The lists consist of transitions with $E_u$/\textit{k} $<$ 300 K and predicted uncertainties lower than 50 kHz, resulting in lines between 0.2 and 300 GHz. 
All of these rest frequencies have at least a precision of 6 $\times$ 10$^{-8}$, corresponding to a radial velocity of 0.018 km s$^{-1}$ or even better. Partition functions at various temperatures for both isomers are listed in Table 2.

\begin{table} [h]
\centering
\caption{Partition functions of $trans$-HCSSH and $cis$-HCSSH at various temperatures.}
\label{table_all}
\setlength{\tabcolsep}{12pt}
\begin{tabular}{c d{-1} d{-1}}
\hline \hline \\ [-2.1ex]
T (K) & \multicolumn{1}{c}{Q ($trans$-HCSSH)} & \multicolumn{1}{c}{Q ($cis$-HCSSH)} \\
\hline \\ [-2ex]
300 & 37126.4660 & 36887.4922
\\ [1ex]
200 & 20421.8656 & 20276.8962
 \\ [1ex]
100 & 7224.8408 & 7172.3000
\\ [1ex]
50  & 2554.3700 & 2535.7348        
\\ [1ex]
20  & 647.0112 & 642.2926
\\ [1ex]
10  & 229.3211 & 227.6550
\\
\hline
\end{tabular}
\end{table}

\section{Discussion and conclusions} \label{sec:disc}
\indent\indent

This laboratory study provides a comprehensive spectroscopic characterization of the two stable conformers of HCSSH in the ground state. Extension of the measurements up to 478 GHz and the resulting larger dataset lead to an overall improvement of the spectroscopic parameters with respect to that of \cite{bak1978}. In particular, the uncertainties on the $A$ rotational constant and on the $D_{J}$ centrifugal distortion constant were reduced by more than 3 orders of magnitude. 
We extended the centrifugal distortion analysis with all the quartic and four sextic constants. Additionally, the octic $L_{KKJ}$ parameter was included for the $trans$ conformer. Further improvements in the fit of the $cis$ conformer were obtained through measurements of $b$-type transitions. 
These new sets of spectroscopic parameters made it possible to compute accurate rest frequencies. In comparison to previous studies, predicted uncertainties were significantly reduced and have values as low as 2 kHz (0.006 km s$^{-1}$) at 3 mm wavelength. 

\begin{figure*} [h]
   \centering
    \includegraphics[width=18.cm]{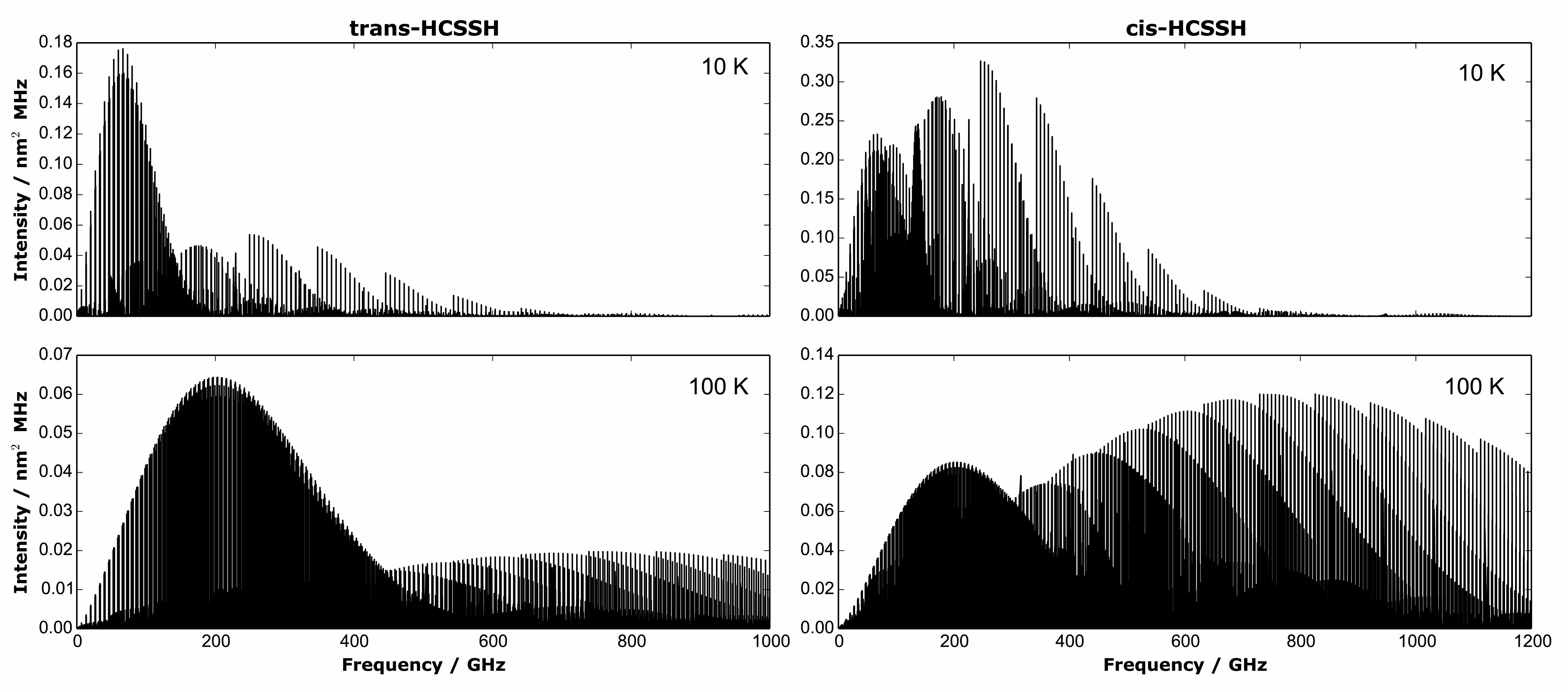}
     \caption{Intensities and line distributions of rotational transitions at 10 and 100 K, for $trans$ and $cis$ dithioformic acid isomers. Most of the bandheads above 250 GHz are produced by $b$-type transitions.}
      \label{spectra}
\end{figure*}

An astronomical search of these molecules can be carried out in the cm, mm, and submm bands, depending on the temperature of the source. As shown in Figure 3, in cold environments with $T_{\mathrm{kin}}\sim$ 10 K or lower (such as the S-rich prestellar cores L183 and Barnard 1), assuming local thermal equilibrium, calculated line intensities and distribution for the $trans$ isomer peak around 45 GHz. In warm regions with $T_{\mathrm{kin}}\sim$ 100 K, the peak shifts at higher frequencies, ca. 200 GHz.
Owing to its larger $b$ component of the dipole moment, the $cis$ isomer shows a different intensity distribution. In this case the calculated peak lies at $\sim$ 280 GHz in cold sources and at $\sim$ 780 GHz in warm regions. However, the strongest lines above 300 GHz, i.e. $b$-type transitions with low $E_u$/\textit{k}, present uncertainties higher than 1 MHz. These large errors are a consequence of the restricted number of measurable $b$-type transitions, strongly limited by the overall lower intensity of the $cis$' lines with respect to the $trans$.       

Work is also under way in our group to extend astrochemical modelling of sulphur to a number of new species, including HCSSH (Laas et al., in prep.). Although this molecule has not been previously studied in the context of astrochemical modelling, a number of reactions may be readily tested, based on standard grain surface rates \citep{Hasegwa1992} or ion-neutral gas-phase kinetics \citep{troe1985}.
Analogous to many other interstellar complex organic molecules, it is likely that a series of barrierless radical-radical addition reactions might lead to its formation on grain surfaces, such as
\begin{equation}
\textrm{HS + CS $\to$ CSSH,}
\end{equation}
\begin{equation}
\textrm{CSSH + H $\to$ HCSSH,}
\end{equation}   
in analogy with the efficient formation of the O-bearing counterpart formic acid. In the gas phase, HCOOH may be efficiently formed via the ionic route,
\begin{equation}  
\begin{tikzcd}[column sep=large,row sep=0.2ex]
& \textrm{H$_3$O$^+$ + CO} \\
\textrm{H$_2$O + HCO$^+$} \arrow[rounded corners,to path={-- ([xshift=2ex]\tikztostart.east) 
                                                          |- (\tikztotarget) [near end,swap]\tikztonodes}]{ur}{>99\%} 
                          \arrow[rounded corners,to path={-- ([xshift=2ex]\tikztostart.east) 
                                                          |- (\tikztotarget) [near end]\tikztonodes}]{dr}[near end]{<1\%} & \\
     & \textrm{(HCOOH)H$^+$}
\end{tikzcd}
\end{equation}
\begin{equation}
\textrm{(HCOOH)H$^+$ + e$^-$ $\to$ HCOOH + H,}
\end{equation}
and the analogue of this route will also be tested for the gas-phase formation of HCSSH. If HCSSH is indeed found to be formed efficiently in a way similar to HCOOH, it is likely that the most promising sources for detection are S-rich regions, where formic acid can be routinely observed.

\begin{acknowledgement} \label{sec:ack}
The authors wish to thank L. Dore for providing the \texttt{proFFit} code.
\end{acknowledgement}


\begin{thebibliography}{57}
\expandafter\ifx\csname natexlab\endcsname\relax\def\natexlab#1{#1}\fi

\bibitem[{{Bak} {et~al.}(1979){Bak}, {Nielsen}, {Svanholt}, \&
  {Christiansen}}]{bak1979}
{Bak}, B., {Nielsen}, O., {Svanholt}, H., \& {Christiansen}, J.~J. 1979, J.
  Mol. Spectr., 75, 134

\bibitem[{{Bak} {et~al.}(1978){Bak}, {Nielsen}, \& {Svanholt}}]{bak1978}
{Bak}, B., {Nielsen}, O.~J., \& {Svanholt}, H. 1978, J. Mol. Spectr., 69, 401

\bibitem[{{Bizzocchi} {et~al.}(2017){Bizzocchi}, {Lattanzi}, {Laas},
  {Spezzano}, {Giuliano}, {Prudenzano}, {Endres}, {Sipil{\"a}}, \&
  {Caselli}}]{bizzocchi2017}
{Bizzocchi}, L., {Lattanzi}, V., {Laas}, J., {et~al.} 2017, \aap, 602, A34

\bibitem[{Calmonte {et~al.}(2016)Calmonte, Altwegg, Balsiger, Berthelier,
  Bieler, Cessateur, Dhooghe, Van~Dishoeck, Fiethe, Fuselier,
  {et~al.}}]{calmonte2016}
Calmonte, U., Altwegg, K., Balsiger, H., {et~al.} 2016, \mnras, 462, S253

\bibitem[{Darling \& Dennison(1940)}]{darling1940}
Darling, B.~T. \& Dennison, D.~M. 1940, Phys. Rev., 57, 128

\bibitem[{Davis {et~al.}(1980)Davis, Robiette, Gerry, Bjarnov, \&
  Winnewisser}]{davis1980}
Davis, R.~W., Robiette, A., Gerry, M., Bjarnov, E., \& Winnewisser, G. 1980, J.
  Mol. Spectr., 81, 93

\bibitem[{Deegan \& Knowles(1994)}]{deegan1994}
Deegan, M.~J. \& Knowles, P.~J. 1994, Chem. Phys. Lett., 227, 321

\bibitem[{Dore(2003)}]{dore2003}
Dore, L. 2003, J. Mol. Spectr., 221, 93

\bibitem[{{Druard} \& {Wakelam}(2012)}]{druard2012}
{Druard}, C. \& {Wakelam}, V. 2012, \mnras, 426, 354

\bibitem[{Dunning~Jr(1989)}]{dunning1989}
Dunning~Jr, T.~H. 1989, J. Chem. Phys., 90, 1007

\bibitem[{Dunning~Jr {et~al.}(2001)Dunning~Jr, Peterson, \&
  Wilson}]{dunning2001}
Dunning~Jr, T.~H., Peterson, K.~A., \& Wilson, A.~K. 2001, J. Chem. Phys., 114,
  9244

\bibitem[{Feller(1992)}]{feller1992}
Feller, D. 1992, J. Chem. Phys., 96, 6104

\bibitem[{Feller(1993)}]{feller1993}
Feller, D. 1993, J. Chem. Phys., 98, 7059

\bibitem[{{Ferrante} {et~al.}(2008){Ferrante}, {Moore}, {Spiliotis}, \&
  {Hudson}}]{ferrante2008}
{Ferrante}, R.~F., {Moore}, M.~H., {Spiliotis}, M.~M., \& {Hudson}, R.~L. 2008,
  \apj, 684, 1210

\bibitem[{{Garozzo} {et~al.}(2010){Garozzo}, {Fulvio}, {Kanuchova}, {Palumbo},
  \& {Strazzulla}}]{garozzo2010}
{Garozzo}, M., {Fulvio}, D., {Kanuchova}, Z., {Palumbo}, M.~E., \&
  {Strazzulla}, G. 2010, \aap, 509, A67

\bibitem[{Gordy \& Cook(1984)}]{gordy1984}
Gordy, W. \& Cook, R.~L. 1984, Microwave molecular spectra (Wiley,)

\bibitem[{Halkier {et~al.}(1999)Halkier, Klopper, Helgaker, \&
  Jo/rgensen}]{halkier1999}
Halkier, A., Klopper, W., Helgaker, T., \& Jo/rgensen, P. 1999, J. Chem. Phys.,
  111, 4424

\bibitem[{Hampel {et~al.}(1992)Hampel, Peterson, \& Werner}]{hampel1992}
Hampel, C., Peterson, K.~A., \& Werner, H.-J. 1992, Chem. Phys. Lett., 190, 1

\bibitem[{{Hasegawa} {et~al.}(1992){Hasegawa}, {Herbst}, \&
  {Leung}}]{Hasegwa1992}
{Hasegawa}, T.~I., {Herbst}, E., \& {Leung}, C.~M. 1992, \apjs, 82, 167

\bibitem[{Heckert {et~al.}(2006)Heckert, K{\'a}llay, Tew, Klopper, \&
  Gauss}]{heckert2006}
Heckert, M., K{\'a}llay, M., Tew, D.~P., Klopper, W., \& Gauss, J. 2006, J.
  Chem. Phys., 125, 044108

\bibitem[{Hehre(1986)}]{warren1986}
Hehre, W.~J. 1986, Ab initio molecular orbital theory (Wiley-Interscience)

\bibitem[{Helgaker {et~al.}(1997)Helgaker, Klopper, Koch, \&
  Noga}]{helgaker1997}
Helgaker, T., Klopper, W., Koch, H., \& Noga, J. 1997, J. Chem. Phys., 106,
  9639

\bibitem[{{Irvine} {et~al.}(1990){Irvine}, {Friberg}, {Kaifu}, {Matthews},
  {Minh}, {Ohishi}, \& {Ishikawa}}]{irvine1990}
{Irvine}, W.~M., {Friberg}, P., {Kaifu}, N., {et~al.} 1990, \aap, 229, L9

\bibitem[{{Jackson} {et~al.}(2004){Jackson}, {Scodinu}, {Xu}, \&
  {Cochran}}]{jackson2004}
{Jackson}, W.~M., {Scodinu}, A., {Xu}, D., \& {Cochran}, A.~L. 2004, \apjl,
  607, L139

\bibitem[{{Jefferts} {et~al.}(1971){Jefferts}, {Penzias}, {Wilson}, \&
  {Solomon}}]{jefferts1971ocs}
{Jefferts}, K.~B., {Penzias}, A.~A., {Wilson}, R.~W., \& {Solomon}, P.~M. 1971,
  \apjl, 168, L111

\bibitem[{Jenkins(1987)}]{jenkins1987}
Jenkins, E.~B. 1987, in Interstellar Processes (Springer), 533--559

\bibitem[{{Jim{\'e}nez-Escobar} \& {Mu{\~n}oz Caro}(2011)}]{jimenezescobar2011}
{Jim{\'e}nez-Escobar}, A. \& {Mu{\~n}oz Caro}, G.~M. 2011, \aap, 536, A91

\bibitem[{{Jim{\'e}nez-Escobar} {et~al.}(2014){Jim{\'e}nez-Escobar}, {Mu{\~n}oz
  Caro}, \& {Chen}}]{jimenezescobar2014}
{Jim{\'e}nez-Escobar}, A., {Mu{\~n}oz Caro}, G.~M., \& {Chen}, Y.-J. 2014,
  \mnras, 443, 343

\bibitem[{Kendall {et~al.}(1992)Kendall, Dunning~Jr, \& Harrison}]{kendall1992}
Kendall, R.~A., Dunning~Jr, T.~H., \& Harrison, R.~J. 1992, J. Chem. Phys., 96,
  6796

\bibitem[{Kolesnikov{\'a} {et~al.}(2014)Kolesnikov{\'a}, Tercero, Cernicharo,
  Alonso, Daly, Gordon, \& Shipman}]{kolesnikova2014}
Kolesnikov{\'a}, L., Tercero, B., Cernicharo, J., {et~al.} 2014, \apjl, 784, L7

\bibitem[{Krishnan \& Pople(1978)}]{krishnan1978}
Krishnan, R. \& Pople, J.~A. 1978, Int. J. Quantum Chem., 14, 91

\bibitem[{{Linke} {et~al.}(1979){Linke}, {Frerking}, \& {Thaddeus}}]{linke1979}
{Linke}, R.~A., {Frerking}, M.~A., \& {Thaddeus}, P. 1979, \apjl, 234, L139

\bibitem[{M{\o}ller \& Plesset(1934)}]{moller1934}
M{\o}ller, C. \& Plesset, M.~S. 1934, Phys. Rev., 46, 618

\bibitem[{Nguyen {et~al.}(1999)Nguyen, Nguyen, \& Le}]{nguyen1999}
Nguyen, M.~T., Nguyen, T.~L., \& Le, H.~T. 1999, J. Phys. Chem. A, 103, 5758

\bibitem[{{Oppenheimer} \& {Dalgarno}(1974)}]{oppenheimer1974}
{Oppenheimer}, M. \& {Dalgarno}, A. 1974, \apj, 187, 231

\bibitem[{{Penzias} {et~al.}(1971){Penzias}, {Solomon}, {Wilson}, \&
  {Jefferts}}]{penzias1971cs}
{Penzias}, A.~A., {Solomon}, P.~M., {Wilson}, R.~W., \& {Jefferts}, K.~B. 1971,
  \apjl, 168, L53

\bibitem[{Peterson \& Dunning~Jr(2002)}]{peterson2002}
Peterson, K.~A. \& Dunning~Jr, T.~H. 2002, J. Chem. Phys., 117, 10548

\bibitem[{{Pickett}(1991)}]{pickett1991}
{Pickett}, H.~M. 1991, J. Mol. Spectr., 148, 371

\bibitem[{Pople {et~al.}(1976)Pople, Binkley, \& Seeger}]{pople1976}
Pople, J.~A., Binkley, J.~S., \& Seeger, R. 1976, Int. J. Quantum Chem., 10, 1

\bibitem[{Pople {et~al.}(1987)Pople, Head-Gordon, \& Raghavachari}]{pople1987}
Pople, J.~A., Head-Gordon, M., \& Raghavachari, K. 1987, J. Chem. Phys., 87,
  5968

\bibitem[{Purvis~III \& Bartlett(1982)}]{purvis1982}
Purvis~III, G.~D. \& Bartlett, R.~J. 1982, J. Chem. Phys., 76, 1910

\bibitem[{Raghavachari {et~al.}(1989)Raghavachari, Trucks, Pople, \&
  Head-Gordon}]{raghavachari1989}
Raghavachari, K., Trucks, G.~W., Pople, J.~A., \& Head-Gordon, M. 1989, Chem.
  Phys. Lett., 157, 479

\bibitem[{Ruffle {et~al.}(1999)Ruffle, Hartquist, Caselli, \&
  Williams}]{ruffle1999}
Ruffle, D., Hartquist, T., Caselli, P., \& Williams, D. 1999, \mnras, 306, 691

\bibitem[{Savage \& Sembach(1996)}]{savage1996}
Savage, B.~D. \& Sembach, K.~R. 1996, Annu. Rev. Astron. Astrophys., 34, 279

\bibitem[{{Sinclair} {et~al.}(1973){Sinclair}, {Fourikis}, {Ribes}, {Robinson},
  {Brown}, \& {Godfrey}}]{sinclair1973h2cs}
{Sinclair}, M.~W., {Fourikis}, N., {Ribes}, J.~C., {et~al.} 1973, Aust. J.
  Phys., 26, 85

\bibitem[{{Thaddeus} {et~al.}(1972){Thaddeus}, {Kutner}, {Penzias}, {Wilson},
  \& {Jefferts}}]{thaddeus1972h2s}
{Thaddeus}, P., {Kutner}, M.~L., {Penzias}, A.~A., {Wilson}, R.~W., \&
  {Jefferts}, K.~B. 1972, \apjl, 176, L73

\bibitem[{{Tieftrunk} {et~al.}(1994){Tieftrunk}, {Pineau des Forets},
  {Schilke}, \& {Walmsley}}]{tieftrunk1994}
{Tieftrunk}, A., {Pineau des Forets}, G., {Schilke}, P., \& {Walmsley}, C.~M.
  1994, \aap, 289, 579

\bibitem[{Trambarulo {et~al.}(1958)Trambarulo, Clark, \&
  Hearns}]{trambarulo1958}
Trambarulo, R., Clark, A., \& Hearns, C. 1958, J. Chem. Phys., 28, 736

\bibitem[{Troe(1985)}]{troe1985}
Troe, J. 1985, Chem. Phys. Lett., 122, 425

\bibitem[{Vidal {et~al.}(2017)Vidal, Loison, Jaziri, Ruaud, Gratier, \&
  Wakelam}]{vidal2017}
Vidal, T.~H., Loison, J.-C., Jaziri, A.~Y., {et~al.} 2017, \mnras, 469, 435

\bibitem[{{Wakelam} {et~al.}(2004){Wakelam}, {Caselli}, {Ceccarelli}, {Herbst},
  \& {Castets}}]{wakelamcaselli2004}
{Wakelam}, V., {Caselli}, P., {Ceccarelli}, C., {Herbst}, E., \& {Castets}, A.
  2004, \aap, 422, 159

\bibitem[{{Wakelam} {et~al.}(2005){Wakelam}, {Caselli}, {Ceccarelli}, {Herbst},
  {Mascetti}, \& {Castets}}]{wakelamcaselli2005}
{Wakelam}, V., {Caselli}, P., {Ceccarelli}, C., {et~al.} 2005, in ESA Special
  Publication, Vol. 577, ESA Special Publication, ed. A.~{Wilson}, 435--436

\bibitem[{Wakelam \& Herbst(2008)}]{wakelamherbst2008}
Wakelam, V. \& Herbst, E. 2008, \apj, 680, 371

\bibitem[{Watson(1977)}]{watson1977b}
Watson, J. K.~G. 1977, in Vibrational Spectra and Structure, ed. J.~Durig,
  Vol.~6 (Elsevier, Amsterdam), 1--89

\bibitem[{Winnewisser {et~al.}(2002)Winnewisser, Winnewisser, Stein, Birk,
  Wagner, Winnewisser, Yamada, Belov, \& Baskakov}]{winnewisser2002}
Winnewisser, M., Winnewisser, B.~P., Stein, M., {et~al.} 2002, J. Mol. Spectr.,
  216, 259

\bibitem[{{Woods} {et~al.}(2015){Woods}, {Occhiogrosso}, {Viti}, {Ka{\v
  n}uchov{\'a}}, {Palumbo}, \& {Price}}]{woods2015}
{Woods}, P.~M., {Occhiogrosso}, A., {Viti}, S., {et~al.} 2015, \mnras, 450,
  1256

\bibitem[{{Zuckerman} {et~al.}(1971){Zuckerman}, {Ball}, \&
  {Gottlieb}}]{zuckerman1971}
{Zuckerman}, B., {Ball}, J.~A., \& {Gottlieb}, C.~A. 1971, \apjl, 163, L41

\end{thebibliography}

\appendix
\section{Theoretical calculations} \label{sec:comp}
\indent\indent

Dithioformic acid has been the subject of numerous ab initio calculations in the past in which  structures and conformational behaviour, isomerization, unimolecular rearrangement and decomposition reaction paths have been studied (\citealt{nguyen1999} and references therein). Apart from \cite{nguyen1999}, who made use of QCISD(T) methods \citep{pople1987}, the older studies had been carried out at the HF-SCF \citep{warren1986} or MP2 (\citealt{moller1934}, \citealt{pople1976}, \citealt{krishnan1978}) levels, also making use of small basis sets and thus leading to less accurate results.
In our work, high level quantum-chemical calculations were performed, focussing on an accurate estimation of dipole moments and equilibrium structures.   

Geometry optimization was carried out via the CCSD(T) method (\citealt{purvis1982}, \citealt{raghavachari1989}, \citealt{hampel1992}, \citealt{deegan1994}). Employing the frozen-core approximation, we used the correlation consistent valence basis sets cc-pVnZ (n=T, Q, 5) \citep{dunning1989} for the hydrogen and carbon atoms and the \textit{d}-augmented basis sets cc-pV(n+d)Z \citep{dunning2001} for the sulphur atoms. We also provide an estimate of the equilibrium energies based on a hierarchical sequence of basis sets, using the complete basis set (CBS) limit extrapolation, as described in \cite{heckert2006}. The molecules have  been considered as planar because of the assumption based on the inertial defect (see Section 3).  
The total energies have been obtained as a sum of the Hartree-Fock self-consistent field (HF-SCF) energy, the electron-correlation contribution at the CCSD(T) level, and the core-valence (CV) correlation effect. CBS limit extrapolation has been applied for the first two values as follows:
\begin{equation}
E_{tot} = E_{\infty}^{HF-SCF} + \Delta E_{\infty}^{CCSD(T)} + \Delta E_{CV}.
\end{equation}

The convergence of the Hartree-Fock CBS limit has been evaluated with (\citealt{feller1992}, \citealt{feller1993})
\begin{equation}
E^{HF-SCF}(n) = E_{\infty}^{HF-SCF} + Ae^{-Bn},
\end{equation}
while the function used to extrapolate the value of electron-correlation energy to the CBS limit is \citep{helgaker1997} written as 
\begin{equation}
\Delta E^{CCSD(T)}(n) = \Delta E_{\infty}^{CCSD(T)} + Cn^{-3}.
\end{equation}

The core-valence correction has been calculated and added as follows:
\begin{equation}
\Delta E_{CV}=E_{ae} - E_{fc},
\end{equation}
where both the all electron ($E_{ae}$) and the frozen-core ($E_{fc}$) energies have been calculated using the cc-pwCVQZ basis set \citep{peterson2002}, avoiding correlation of 1s electrons of sulphur in any computation. All the calculations have been performed with the CFOUR program package (www.cfour.de).

\begin{table*} [h]
\centering
\caption{Equilibrium geometries and energies of $trans$-HCSSH and $cis$-HCSSH are reported. All the values were computed at the CCSD(T) level of theory. }
\label{table_all}
\setlength{\tabcolsep}{7pt}
\begin{tabular}{l c c c l c c c c}
\hline \hline \\[-2.3ex]
Basis set & Energy (Hartree) & C=S (\AA) & C-S (\AA) & S-H (\AA) & C-H (\AA) & $\angle$\tablefootmark{a} S=C-S ($^{\circ}$) & $\angle$ C-S-H ($^{\circ}$) & $\angle$ H-C=S ($^{\circ}$) \\
\hline \\ [-2ex]
$trans$-HCSSH &  &  &  &  &  &  &  & \\
\hline \\ [-2ex]
cc-pVTZ       & -834.692586 & 1.6259 & 1.7412 & 1.3418 & 1.0896 & 128.27 & 95.49 & 110.23
\\ [1ex]                      
cc-pVQZ       & -834.739631 & 1.6223 & 1.7352 & 1.3421 & 1.0891 & 128.09 & 95.61 & 110.43
\\ [1ex]                      
cc-pV5Z       & -834.754180 & 1.6212 & 1.7332 & 1.3420 & 1.0891 & 128.04 & 95.64 & 110.50
\\ [1ex]                                           
CBS\tablefootmark{b}        & -834.767911 & 1.6201 & 1.7314 & 1.3420 & 1.0891 & 127.98 & 95.66 & 110.54
\\ [1ex]                      
cc-pwCVQZ(fc) & -834.743529 & 1.6214 & 1.7341 & 1.3415 & 1.0891 & 128.09 & 95.61 & 110.44
\\ [1ex]                      
cc-pwCVQZ(ae) & -835.433398 & 1.6173 & 1.7297 & 1.3394 & 1.0877 & 128.11 & 95.61 & 110.44
\\ [1ex]                      
CBS+CV\tablefootmark{c}    & -835.457780 & 1.6161 & 1.7270 & 1.3398 & 1.0877 & 127.99 & 95.66 & 110.54
\\
\hline \\[-2ex]
$cis$-HCSSH &  & & & & & & &
\\ 
\hline \\ [-2ex]
cc-pVTZ       & -834.690500 & 1.6247 & 1.7454 & 1.3412 & 1.0881 & 123.75 & 96.59 & 114.16
\\ [1ex]                      
cc-pVQZ       & -834.737690 & 1.6211 & 1.7391 & 1.3413 & 1.0877 & 123.58 & 96.79 & 114.33
\\ [1ex]                      
cc-pV5Z       & -834.752258 & 1.6200 & 1.7370 & 1.3412 & 1.0877 & 123.52 & 96.83 & 114.39
\\ [1ex]                                           
CBS\tablefootmark{b}        & -834.765988 & 1.6189 & 1.7352 & 1.3411 & 1.0877 & 123.46 & 96.86 & 114.45
\\ [1ex]                      
cc-pwCVQZ(fc) & -834.741583 & 1.6202 & 1.7380 & 1.3407 & 1.0877 & 123.56 & 96.80 & 114.34
\\ [1ex]                      
cc-pwCVQZ(ae) & -835.431447 & 1.6161 & 1.7336 & 1.3385 & 1.0862 & 123.55 & 96.82 & 114.37
\\ [1ex]                      
CBS+CV\tablefootmark{c}    & -835.455852 & 1.6149 & 1.7309 & 1.3389 & 1.0862 & 123.45 & 96.88 & 114.48
\\
\hline
\end{tabular}
\\ [-1ex]
\tablefoot{
\tablefootmark{a}{Bond angle in degree.} \\
\tablefoottext{b}{CBS extrapolation using the additivity assumption shown in Eq. (A.1) without core-valence correction, $\Delta E_{CV}$.} \\
\tablefoottext{c}{Final best estimated values, calculated using the complete additivity assumption shown in Eq. (A.1).}
}
\end{table*}

The results show that the $trans$ conformer is the most stable and has a difference in energy of 421 cm$^{-1}$ with respect to the $cis$-HCSSH.
This value is in reasonable agreement with the experimental value given by \cite{bak1978}, which differs only by 71 cm$^{-1}$.
Molecular structures were obtained with the same method shown above. All the results are listed in Table A.1.

Dipole moments were also calculated and extrapolated to the CBS limit with a similar procedure \citep{halkier1999}. For these dipole calculations, we used similar basis sets, but augmented with diffuse functions to ensure correct evaluation of this property: that is, the light atoms used the aug-cc-pVnZ basis sets \citep{kendall1992} and the sulphur atoms used the aug-cc-pV(n+d)Z basis sets \citep{dunning2001}. The computations were carried out using the $CBS+CV$ equilibrium geometries, i.e. the best estimated structure obtained with the additivity assumption from Eq.(A.1). The final values are shown in Table A.2.
Owing to its higher computational cost, the barrier energy to $trans-cis$ conversion was computed at the CCSD(T)/cc-pVTZ level, with the frozen-core approximation, giving a value of 3762 cm$^{-1}$.

\begin{table} [h]
\centering
\caption{Dipole moments of $trans$-HCSSH and $cis$-HCSSH. All the values were computed at the CCSD(T) level of theory.}
\label{table_all}
\setlength{\tabcolsep}{12pt}
\begin{tabular}{l c c}
\hline \hline \\[-2.1ex]
Basis set & $\mu_a$ (Debye) & $\mu_b$ (Debye) \\
\hline \\ [-2ex]
$trans$-HCSSH &  & \\
\hline \\ [-2ex]
aug-cc-pVTZ       & 1.4788 & 0.1950
\\ [1ex]
aug-cc-pVQZ       & 1.4870 & 0.1957
 \\ [1ex]
aug-cc-pV5Z       & 1.4807 & 0.1939
\\ [1ex]
CBS\tablefootmark{a}           & 1.4800 & 0.1935        
\\ [1ex]
aug-cc-pCVQZ(fc) & 1.4863 & 0.1945
\\ [1ex]
aug-cc-pCVQZ(ae) & 1.4822 & 0.1934
\\ [1ex]
CBS+CV\tablefootmark{b}        & 1.4766 & 0.1924
\\
\hline \\ [-2ex]
$cis$-HCSSH &  & 
\\ 
\hline \\ [-2ex]
aug-cc-pVTZ & 2.0756 & 1.6323
\\ [1ex]
aug-cc-pVQZ & 2.0896 & 1.6474
\\ [1ex]
aug-cc-pV5Z & 2.0878 & 1.6448
\\ [1ex]
CBS\tablefootmark{a}     & 2.0861 & 1.6428
\\ [1ex]
aug-cc-pwCVQZ(fc) & 2.0883 & 1.6489
\\ [1ex]
aug-cc-pwCVQZ(ae) & 2.0834 & 1.6464
\\ [1ex]
CBS+CV\tablefootmark{b}    & 2.0829 & 1.6403
\\
\hline
\end{tabular}
\\ [-1ex]
\tablefoot{
\tablefoottext{a}{CBS extrapolation using the additivity assumption shown in Eq. (A.1) without core-valence correction,  $\Delta E_{CV}$.} \\
\tablefoottext{b}{Final best estimated values, calculated using the complete additivity assumption shown in Eq. (A.1).}
}
\end{table}

\end{document}